\documentclass[aps, twocolumn, superscriptaddress, showpacs, floatfix]{revtex4-1}
\usepackage{amsmath}
\usepackage{hyperref}
\usepackage{graphicx}
\usepackage{float}
\usepackage{multirow}
\usepackage{color}
\usepackage{soul}

\hypersetup{
  colorlinks = true, 
  urlcolor   = blue, 
  linkcolor  = blue, 
  citecolor  = blue  
}

\begin{document}
\title{Mott transition in the  triangular lattice Hubbard model:\\ a dynamical cluster approximation study}

\author{Hung T. Dang}
\affiliation{Institute for Theoretical Solid State Physics, JARA-FIT and JARA-HPC, RWTH Aachen University, 52056 Aachen, Germany}
\author{Xiao Yan Xu}
\affiliation{Beijing National Laboratory for Condensed Matter Physics,
and Institute of Physics, Chinese Academy of Sciences, Beijing 100190, China}
\author{Kuang-Shing Chen}
\affiliation{Institut f\"ur Theoretische Physik und Astrophysik, Universit\"at W\"urzburg, Am Hubland, D-97074 W\"urzburg, Germany}
\author{Zi Yang Meng}
\affiliation{Beijing National Laboratory for Condensed Matter Physics,
and Institute of Physics, Chinese Academy of Sciences, Beijing 100190, China}
\author{Stefan Wessel}
\affiliation{Institute for Theoretical Solid State Physics, JARA-FIT and JARA-HPC, RWTH Aachen University, 52056 Aachen, Germany}
\date{\today}

\begin{abstract}
Based on  dynamical cluster approximation (DCA) quantum Monte Carlo simulations, we study the interaction-driven Mott metal-insulator transition (MIT) in the half-filled Hubbard model on the anisotropic two-dimensional triangular lattice, where the degree of frustration is varied between the unfrustrated case and the fully frustrated, isotropic triangular lattice. Upon increasing the DCA cluster size, we analyze the evolution of the MIT phase boundary as a function of frustration in the phase diagram spanned by the interaction strength and temperature, and provide a quantitative description of the MIT phase boundary in the triangular lattice Hubbard model. Qualitative differences in the phase boundary between the unfrustrated and fully frustrated cases are exhibited. In particular, a change in the sign of the phase boundary slope is observed, which via an impurity cluster eigenstate analysis, may be related to a change in the nature of the insulating state. We discuss our findings within the scenario that the triangular lattice electron system might exhibit a quantum critical Mott MIT with a possible quantum spin liquid insulating state, such as considered for the organic charge transfer salts $\kappa$-(BEDT-TTF)$_{2}$Cu$_{2}$(CN)$_{3}$ and EtMe$_{3}$Sb[Pd(dmit)$_{2}$]$_{2}$.
\end{abstract}


\maketitle

\section{Introduction\label{sec:intro}}
The Mott metal-insulator transition (MIT)~\cite{Imada1998,mott1990metal} is a fundamental phenomenon in the field of strongly correlated quantum many-body physics and quantum criticality~\cite{Sondhi97,sachdev2001quantum}. Starting from the early case of NiO~\cite{Mott1937}, the full Mott MIT (temperature versus pressure) phase diagram has later been constructed e.g. for vanadium oxide V$_2$O$_3$ \cite{McWhan1973}, and critical scaling near the end point of the Mott MIT phase boundary has been observed in conductance measurements~\cite{Limelette03102003}.
 
The discovery of strongly correlated organic materials opened new horizons for exploring the Mott MIT. These compounds consist of quasi-two-dimensional triangular lattices of organic charge transfer salts~\cite{Shimizu2003,SYamashita2008,MYamashita2008,Itou2008, SYamashita2011,Kanoda2011} such as $\kappa$-(BEDT-TTF)$_{2}$Cu$_{2}$(CN)$_{3}$ and EtMe$_{3}$Sb[Pd(dmit)$_{2}$]$_{2}$. By varying hydrostatic pressure, these systems exhibit a MIT with an insulating state without any detectable symmetry breaking, possibly exhibiting a quantum spin liquid state \cite{Balents2010}. The phase diagram of $\kappa$-(BEDT-TTF)$_{2}$Cu$_{2}$(CN)$_{3}$ in addition exhibits a  superconductivity regime separating the Fermi liquid and the Mott insulating phase \cite{Kurosaki2005}. The interplay between geometric frustration and strong electronic interactions is considered crucial for the possibility of the quantum spin liquid phase and unconventional critical behavior exhibited at the MIT itself~\cite{Kagawa2005}.

On the theoretical side, dynamical mean-field theory (DMFT)~\cite{Georges1996} and its cluster extensions~\cite{Maier2005,Kotliar2006} have been established as powerful tools for the study of strongly correlated materials. Over the past decades, DMFT has been used intensively to investigate the Hubbard model~\cite{Hubbard1963}, the ``simplest nontrivial'' model for correlated electrons, which is in fact a prominent low-energy model for the organic charge transfer salts \cite{Powell2011}. Along with other numerical approaches, DMFT has revealed the MIT phase diagram of the unfrustrated square lattice Hubbard model at half-filling~\cite{Georges1996,Park2008,Gull2008,Werner2009b,Gull2009,Schafer2014}, as well as of extended Hubbard models for multi-orbital systems, resembling transition-metal oxides~\cite{Chan2009,Werner2009a,Huang2012,Georges2013,Du2013,Meng2014a}. 
DMFT studies have furthermore been performed for electron systems with a frustrated lattice geometry. Results for diagonally frustrated two-dimensional square as well as triangular lattices have been obtained from DMFT studies with exact diagonalization~\cite{Kyung2006a,Liebsch2009}, noncrossing approximation \cite{Galanakis2009} , Hirsch-Fye quantum Monte Carlo (QMC)~\cite{Ohashi2008}, or continuous-time QMC cluster impurity solvers~\cite{HunpyoLee2008,Sato2012a,Chen2013a}. With frustration, not only is the position of the van-Hove singularity shifted, and thus antiferromagnetic fluctuations are suppressed at half-filling, several other important aspects  such as the re-entrance to insulating behavior \cite{Limelette03,Kanoda2011,Ohashi2008} or unconventional superconductivity upon hole doping \cite{Chen2013a,Chen2013b} have been discussed.
However, due to a severe minus-sign problem in the QMC simulations, a systematic cluster DMFT study of the evolution of the MIT phase diagram from the unfrustrated limit to the strongly frustrated  triangular lattice is still lacking, especially for the evolution of the MIT phase boundary upon increasing the  cluster size at low temperature, which may encode changes in the MIT physics. 

Here, we aim at providing such a systematic study by employing cluster DMFT simulations based on the dynamical cluster approximation (DCA) with continuous-time QMC as cluster solvers~\cite{Gull2011} to analyze the evolution of the MIT phase boundary  in the anisotropic two-dimensional triangular lattice Hubbard model. We focus on the low-temperature regime, for which our results reveal a qualitative difference in the phase boundary between the unfrustrated and the fully frustrated triangular lattice, suggesting a connection between the MIT on the  frustrated triangular lattice and the possible emergence of a spin liquid phase in a regime, where conventional magnetic order is suppressed by the interplay of geometric frustration and electronic correlations.

The rest of the paper is organized as follows. Section~\ref{sec:model_methods} introduces the model and numerical methods employed in this study. In Sec.~\ref{sec:phase_diagrams}, the MIT phase boundaries for different degrees of frustration and cluster sizes are presented. In Sec.~\ref{sec:discussion}, 
we discuss the changes in the phase boundary as the degree of frustration increases from several physical perspectives. Section~\ref{sec:conclusions} contains our conclusion and outlook. Finally, an  appendix provides quantitative details on  the sign problem in our QMC simulations.

\section{Model and Methods\label{sec:model_methods}}
\subsection{Model}
\begin{figure}
\includegraphics[width=\columnwidth]{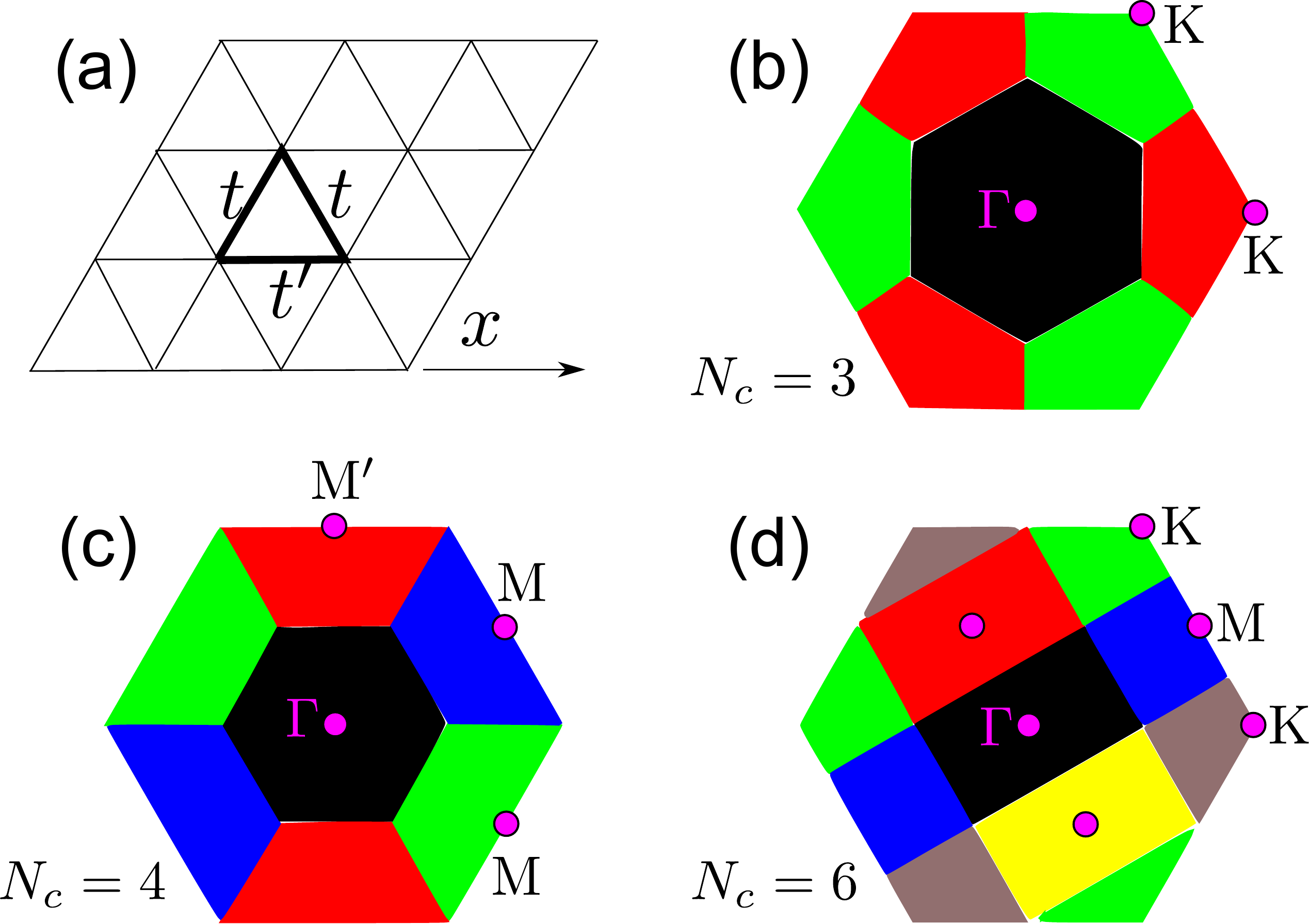}
\caption{(Color online) (a): Anisotropic triangular lattice with nearest neighbor hopping $t'$ along $x$ direction and $t$ along the other directions. (b)-(d): Brillouin zone (BZ) and momentum patches with $N_{c}=3$, $4$, and $6$ clusters employed in the  DCA simulations. Each patch is represented by the corresponding cluster momentum (solid dots - magenta online).}
\label{fig:triangular_lattice}
\end{figure}

We consider the half-filled one-band Hubbard model on the triangular lattice, described by the Hamiltonian
\begin{equation}\label{eq:Hubbard}
 H = -\sum_{\langle ij\rangle\sigma} t_{ij} c^\dagger_{i\sigma} c_{j\sigma} + U\sum_i n_{i\uparrow} n_{i\downarrow},
\end{equation}
where $\langle ij\rangle$ denotes nearest neighbor sites and the hopping parameter $t_{ij}$ equals either $t$ or $t'$, as sketched in Fig.~\ref{fig:triangular_lattice}(a), with the hopping $t'$ along the $x$-direction varied from $0$ to $t$. The unfrustrated limit $t'=0$ is topologically equivalent to a square lattice, while $t'=t$ is the fully frustrated triangular lattice case. $U$ is the on-site Coulomb repulsion, which is varied across the MIT boundary.

\subsection{Dynamical cluster approximation}
To treat strong correlation effects, we employ the DCA~\cite{Maier2005}, a cluster extension of the single-site DMFT \cite{Georges1996}, using  continuous-time QMC as cluster impurity solvers \cite{Gull2011}. Within this approach, the original lattice model is mapped onto a periodic cluster of size $N_c$, dynamically embedded in a self-consistently determined mean-field bath. Spatial correlations inside the cluster are treated explicitly while those at larger distances are described on the mean-field level. Temporal correlations, essential for quantum criticality, are treated explicitly by the QMC solver both on the cluster as well as in the bath.

In DCA, the Brillouin zone (BZ) is divided into $N_c$ momentum patches, each of which is generally denoted as $\mathbf{K}$, as shown in Figs.~\ref{fig:triangular_lattice} (b)-(d) for $N_{c}=3$, $4$, $6$, respectively. The cluster self-energy is used to approximate the lattice self-energy, i.e., $\Sigma(\mathbf{k},\omega) = \Sigma_\mathbf{K}(\omega)$ if $\mathbf{k} \in \mathbf{K}$.

The DCA self-consistency condition is achieved when the cluster Green's function $G_{\mathbf{K}}(i\omega_n)$ is equal to the coarse-grained lattice Green's function, where the latter is obtained from the coarse-graining process with the cluster self-energy $\Sigma_\mathbf{K}(\omega)$ via
\begin{equation}\label{eq:self_consistent_eqn}
 \bar{G}_\mathbf{K}(i\omega_n) = \dfrac{N_c}{N} \sum_{\mathbf{k} \in \mathbf{K}} \dfrac{1}{i\omega_n +\mu - \epsilon_\mathbf{k} - \Sigma_\mathbf{K}(i\omega_n)},
\end{equation}
and the bare dispersion for the anisotropic triangular lattice is given by 
\begin{equation}\label{eq:dispersion}
 \epsilon_\mathbf{k} = -2t' \cos(k_x) - 4 t \cos \left(\dfrac{\sqrt{3}}{2}k_y\right) \cos\left(\dfrac{k_x}{2}\right).
\end{equation}

An arbitrary initial guess for the bare Green's function, usually the non-interacting one, is taken as the DCA input for the cluster impurity solver. The cluster self-energy is obtained after solving the impurity model, containing information about the correlations. It is then used to compute $\bar{G}_\mathbf{K}(i\omega_n)$ based on Eq.~\eqref{eq:self_consistent_eqn}. The bare Green's function serving as the input for the next iteration is obtained via the Dyson equation as $(\bar{G}^{-1}_\mathbf{K}(i\omega_n)+\Sigma_{\mathbf{K}}(i\omega_n))^{-1}$. Once convergence of the self-consistent calculation is achieved, a final iteration is necessary to measure physical quantities.

The bottleneck of the  DCA calculations resides in solving the cluster problem. Depending on the employed cluster solver, the computational effort scales either exponentially with the cluster size or to the cube of the inverse temperature times cluster size when the sign problem is absent~\cite{Gull2011}. For frustrated systems, the QMC minus-sign problem makes the calculations challenging [see Appendix~\ref{app:sign_problem}]. To obtain good-quality results, we employ two different continuous-time QMC cluster solvers in a complementary manner: for low temperatures we employ the hybridization expansion solver [CT-HYB]~\cite{Werner2006,Werner2006a} while for large cluster sizes we use the interaction expansion solver [CT-INT]~\cite{Rubtsov2005}. In particular, CT-HYB is used for  $N_c=1, 3$ and $4$ DCA simulations, for temperatures as low as $T=0.025t$ and $t'$ varying from 0 to $t$. CT-INT is used for $N_c=6$ DCA simulations for the isotropic triangular lattice $(t'=t)$, as well as for the unfrustrated case $(t'=0)$ with $N_c=16$, extending thus beyond previous low-temperature results \cite{Gull2009} with $N_c=8$ for the square lattice. 

\subsection{Total energy and entropy}
Following Refs.~\cite{Georges1996,Werner2005,Mikelsons2009,Fuchs2011,LeBlanc2013}, we extract the system's entropy (per site) based on a thermodynamic integration of the total energy (per site), $E(\beta)$, where $\beta=1/T$ denotes the inverse temperature. 
At half-filling and for $\beta=0$, the entropy per site equals $S(\beta=0)=\ln 4$ \cite{Georges1996}. For $\beta > 0$, the entropy is then calculated as
\begin{equation}\label{eq:entropy}
 S(\beta) = S(\beta=0) + \beta E(\beta) - \int_0^\beta E(\beta') d\beta'.
\end{equation}
The total energy $E(\beta)$  is obtained from the DCA measurements as follows: the kinetic energy is $\sum_{\langle ij\rangle\sigma} t_{ij}\langle c^\dagger_{i\sigma}c_{j\sigma}\rangle = \sum_{\mathbf{k}\sigma} \epsilon_\mathbf{k}\langle c^\dagger_{\mathbf{k}\sigma}c_{\mathbf{k}\sigma}\rangle$, where 
\[
\langle c^\dagger_{\mathbf{k}\sigma}c_{\mathbf{k}\sigma}\rangle = \dfrac{1}{\beta} \sum_{i\omega_n} 
  \dfrac{e^{i\omega_n 0^+}}{i\omega_n + \mu - \epsilon_\mathbf{k} - \Sigma_\mathbf{K}(i\omega_n)},
\]
while the interaction energy $U\sum_i \langle n_{i\uparrow} n_{i\downarrow}\rangle$ is calculated from the cluster double occupancy.
The accuracy of the entropy obtained using Eq.~\eqref{eq:entropy} depends on the chosen reference entropy \cite{LeBlanc2013}, which in our case is taken as $S(\beta=0)$. For different reference values, the entropy can be shifted by constants. Details of the entropy calculation and a comparison with the results in Ref.~\onlinecite{Kokalj2013}, based on finite temperature Lanczos calculations,  are presented in Appendix~\ref{app:entropy}.

\subsection{Determining the metallic/insulating state}
We prefer to study the MIT based on directly accessible Matsubara frequency DCA data.
The Mott MIT can be determined from the Matsubara Green's function by considering the expression
\begin{equation}\label{eq:imag_green}
\begin{split}
G(i\omega_n) &= \int \dfrac{A(\omega)d\omega}{i\omega_n - \omega} \\
             &\approx -P \int\dfrac{A(\omega)d\omega}{\omega} - i\pi A(\omega=0),
\end{split}
\end{equation}
where $P$ denotes the principle value. In this expression, $\omega_n$ is assumed to be small so that the Sokhotski-Plemelj identity $1/(x-i\gamma)=P (1/x) + i\pi \delta(x)$ (where $\gamma$ is an infinitesimal quantity) applies. Hence, the spectral weight for each momentum sector $\mathbf{K}$ at the Fermi level is given as $A_{\mathbf{K}}(\omega=0) \approx -\mathrm{Im}~G_{\mathbf{K}}(i\omega_0)/\pi$, with $\omega_0 = \pi T$ the lowest Matsubara frequency available at temperature $T$.
\begin{figure}
 \includegraphics[width=\columnwidth]{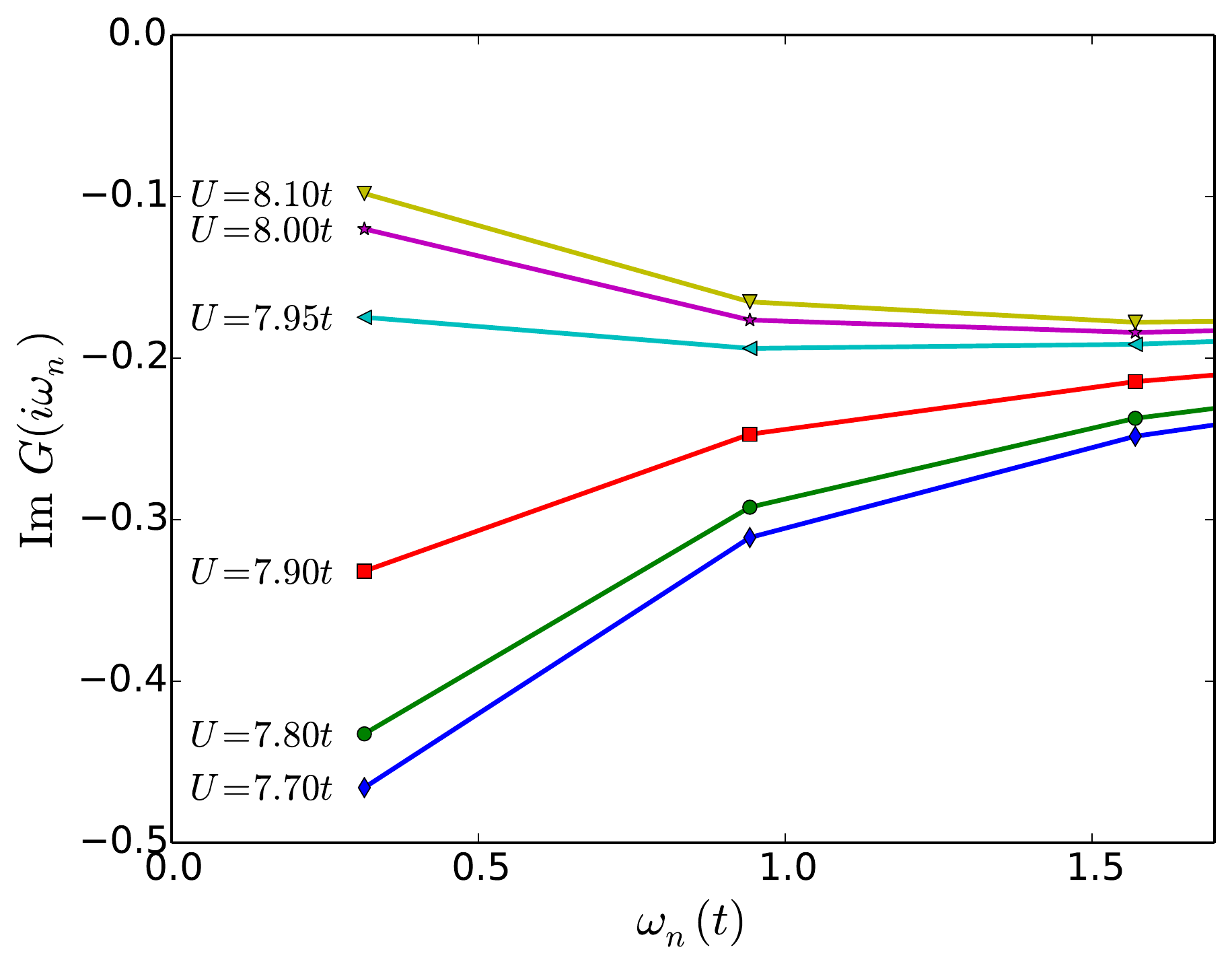}
\caption{\label{fig:imag_green} (Color online) Imaginary part of the Green's function in the $\mathbf{K}=M$ sector [cf. Fig.~\ref{fig:triangular_lattice}(c)] at the three lowest Matsubara frequencies, calculated for $t'=0.9t$ at $T=0.1t$ within $N_c=4$ DCA. The value of $U_c$ for the MIT is estimated between $7.90t$ and $7.95t$.}
\end{figure} 
Thus, in order to assess, if an energy gap opens at the Fermi level, one may consult the value of $\mathrm{Im}~G(i\omega_0)$. However, since it is difficult to draw conclusions from a single data point, in practice we consider $\mathrm{Im}~G(i\omega_n)$ at several low Matsubara frequencies and then extract the low-frequency tendency. As demonstrated in Fig.~\ref{fig:imag_green}, a momentum sector is in a metallic state if $\mathrm{Im}~G(i\omega_n)$ approaches a finite value as $\omega_n\to 0$, whereas if it bends to zero, the sector is insulating. The critical value $U_c$ is estimated from  the two closest $U$-values at which the sector changes from metallic to insulating behavior. We have verified that this approach, directly based on the DCA measurements,  is consistent with the observation of the spectral function as obtained after an analytic continuation (not shown). The direct approach  performs well at low temperatures such that the approximation in Eq.~\eqref{eq:imag_green} is appropriate. Accordingly, in our study, we mainly focus on the low-temperature region, $T\le 0.1t$.

We note that the Green's function does not usually behave in the same way for different momentum sectors. Therefore the general idea is to check all the momentum sectors for the MIT. However, in practice, the $\Gamma$ sector [see Fig.~\ref{fig:triangular_lattice}] lies completely within the Fermi surface; it is always insulating and thus can be safely disregarded. For $N_c\le 4$, both the $K$ and $M$ sector [Fig.~\ref{fig:triangular_lattice}(b), (c)] contains part of the Fermi surface, they behave similarly and have the same $U_c$ for the MIT, thus we require to consider only one of these momentum sectors. For $N_c\ge 6$, there appears a sector-selective Mott transition \cite{Gull2009}, and therefore we require to consider all the cluster momentum points close to the Fermi surface [see Fig.~\ref{fig:triangular_lattice}(d)] for the MIT.

\section{Metal-insulator transition and the phase diagrams\label{sec:phase_diagrams}}
In this section, we determine the MIT phase diagrams from DCA calculations upon varying the impurity cluster size $N_c$.
Although differences in the phase boundaries arise for different cluster sizes, a consistent picture of the phase boundary emerges as the cluster size increases. 

\begin{figure}
 \includegraphics[width=\columnwidth]{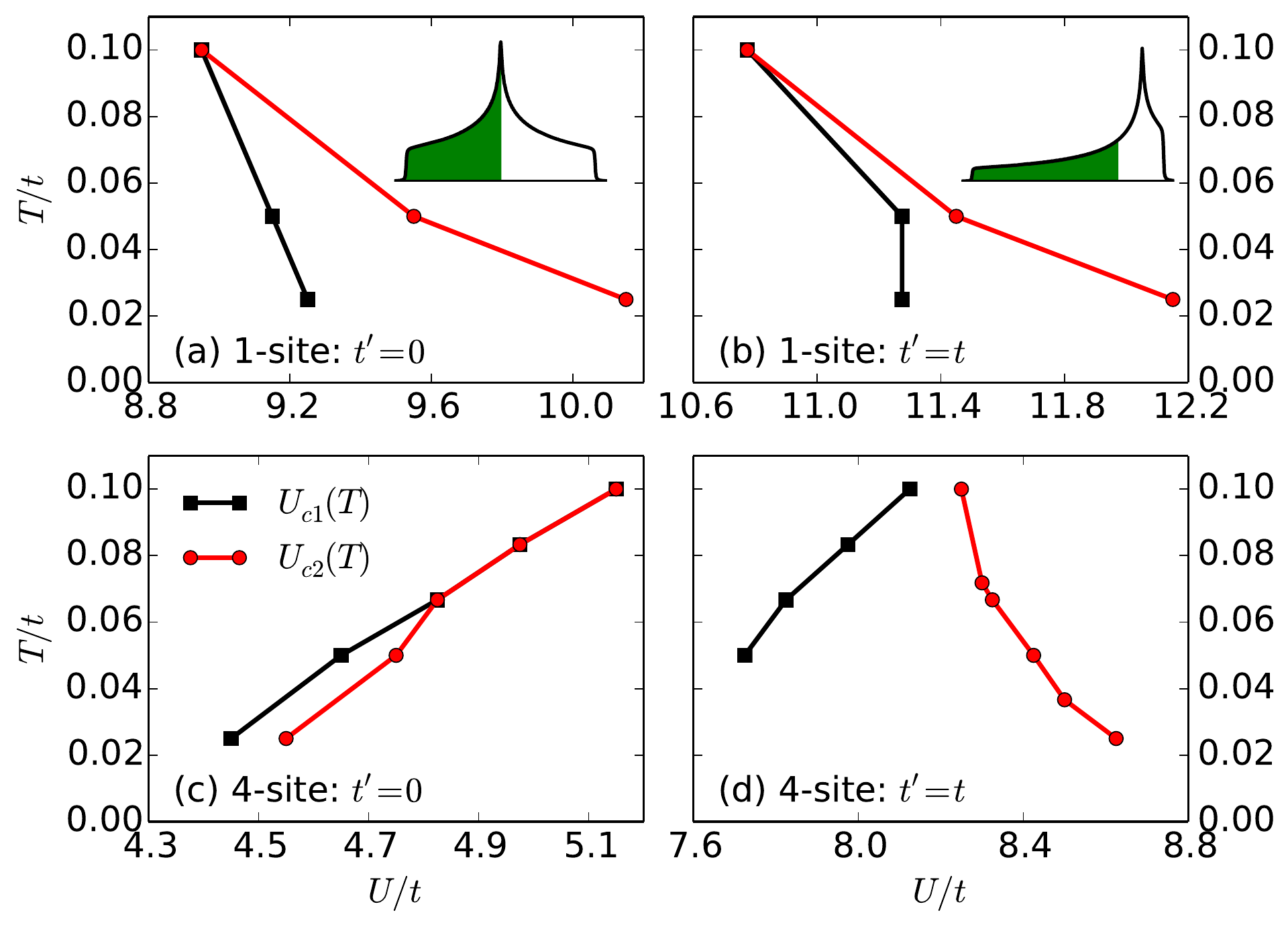}
\caption{\label{fig:phase_diagram_1site_4site} (Color online)  MIT phase diagrams calculated using $N_c=1$ DMFT (the first row) and $N_c=4$ DCA (the second row) for the unfrustrated ($t'=0$) and fully frustrated ($t'=t$) triangular lattice. The coexistence regime is bound by the $U_{c1}(T)$ (black squares) and $U_{c2}(T)$ (red circles) phase boundaries. The leftmost (rightmost) side of the phase diagrams is metallic (insulating). The insets in panels (a) and (b) show the non-interacting system's DOS  filled up to the half-filled level, for $t'=0$ and $t'=t$, respectively.}
\end{figure}

Single-site DMFT is known to be exact in  infinite dimensions \cite{Georges1996} where only local correlations are present. However, for low-dimensional systems such as the triangular lattice ($d=2$) considered in this study, spatial correlation effects are important, and need to be taken into consideration. In Fig.~\ref{fig:phase_diagram_1site_4site}, we show how the shapes of the DCA phase diagrams crucially change when extending beyond single-site DMFT ($N_c=1$) to $N_c=4$ DCA calculations at $t'=0$ and $t'=t$. Similar to Refs.~\cite{Georges1996,Park2008}, for each panel of Fig.~\ref{fig:phase_diagram_1site_4site},  two curves $U_{c1}(T)$ and $U_{c2}(T)$ are shown that separate stable metallic (left) and insulating (right) regions, while in the intermediate coexistence region both metallic and insulating solutions are available, depending on the initial condition chosen when iterating  the self-consistency loop.

When extending beyond $N_c=1$ DMFT to $N_c=4$ DCA, we find qualitative changes both in the shape of the phase boundaries and in the numerical values of the transition lines $U_c(T)$. The effects of non-local correlations, that account for these changes, thus depend strongly on the degree of frustration. Considering  the unfrustrated case, $t'=0$, our results in Figs.~\ref{fig:phase_diagram_1site_4site}(a) and (c) are compatible with previous results in the literature~\cite{Park2008,Gull2008}. The substantial decrease of the  $U_c$ values from $N_c=1$ to $N_c=4$ DCA for the $t'=0$ case relates to the fact that antiferromagnetic (AFM) nesting vectors at the Fermi surface are included in the momentum patches of the $N_c=4$ cluster, such that AFM fluctuations are taken into account, causing a Slater mechanism to open a gap~\cite{Imada1998,Gull2008,Schafer2014}. The change in the slope of the phase boundaries from negative [$N_c=1$ DMFT - Fig.~\ref{fig:phase_diagram_1site_4site}(a)] to positive [$N_c=4$ DCA - Fig.~\ref{fig:phase_diagram_1site_4site}(c)] is related to the change of the entropy in the insulating state: while within $N_c=1$ DMFT free local spins reside on the impurity site,  $N_c=4$ DCA exhibits a cluster singlet state within the insulating phase~\cite{Park2008}. 

Moreover, because of the perfect nesting, the $U_c$ value in the unfrustrated case is known to approach to  $0$ as $N_c\to\infty$ and $T\to 0$, i.e. antiferromagnetic fluctuations open a gap at arbitrarily weak, finite $U>0$~\cite{Hirsch1985,Varney2009,Schafer2014}. This is reflected by a decrease of $U_c$ by a factor of $2$ between the $N_c=1$ and $N_c=4$ calculations [Figs.~\ref{fig:phase_diagram_1site_4site}(a) and (c)]. However, upon going from $N_c=4$ to $8$, $U_c$ increases, which is due to a momentum-sector-selective MIT~\cite{Werner2009b,Gull2009}. For $N_c=8$, at $T=0.05t$, momentum sectors $\mathbf{K}=(0,\pi)$ and $(\pi,0)$ open the gap first at $U_s\approx5.6t$, while sector $\mathbf{K}=(\frac{\pi}{2},\frac{\pi}{2})$ needs a larger $U_{c2}\approx6.4t$ to open the gap, and hence $U_{c2}$ (at which all the momentum sectors are gapped) for $N_c=8$ is larger than the $U_{c2}$ value for $N_c=4$ where there is no sector-selective MIT \cite{Gull2008,Werner2009b,Gull2009}. To clarify the situation, we performed DCA simulations with an even larger cluster size, $N_c=16$, for the $t'=0$ case at low temperature as $T=0.05t$, and obtain $U_s\approx3.75t$ for momentum sectors $\mathbf{K}=(0,\pi)$ and $(\pi,0)$, and $U_{c2}\approx5.2t$ for the momentum sector $\mathbf{K}=(\frac{\pi}{2},\frac{\pi}{2})$. Both values clearly fall below the corresponding values for $N_c=8$~\cite{Gull2009}. We thus believe that, as long as the cluster size $N_c$ is large enough to observe the momentum-sector-selective MIT, the $U_c$ value for the transition decreases monotonically as $N_c$ increases.

The problem becomes more difficult, once frustration is introduced into the system by turning on finite values of $t'$. Figure~\ref{fig:phase_diagram_1site_4site}(b) shows that the phase diagrams for the fully frustrated system ($t'=t$) are already different from the unfrustrated case on  the single-site DMFT level. In particular, the frustrated case exhibits larger values of  $U_{c2}$ (namely, $\sim 10.2t$ for $t'=0$ versus $\sim 12.2t$ for $t'=t$), the $U_{c1}(T)$ curve exhibits a different slope, and the coexistence region is considerately smaller, e.g. less than half the $t'=0$ value at $T=0.05t$. In single-site DMFT, the only contribution to these difference is the shape of the free system's DOS. As shown in the insets of Fig.~\ref{fig:phase_diagram_1site_4site}(a), and (b), the DOS in the $t'=0$ case exhibits a van Hove singularity right at the Fermi level, while it is shifted towards the upper band edge for $t'=t$. In addition,  the electronic bandwidth $W$ in the  frustrated case is slightly larger ($W_{t'=0}=8t$; $W_{t'=t} = 9t$). The  free $t'=0$ electron system is thus more susceptible towards interaction effects, while the $t'=t$ case can reside within a more stable Fermi liquid state. These appear to be the main explanations for the differences seen between the two cases within single-site DMFT calculations.

\begin{figure}
 \includegraphics[width=\columnwidth]{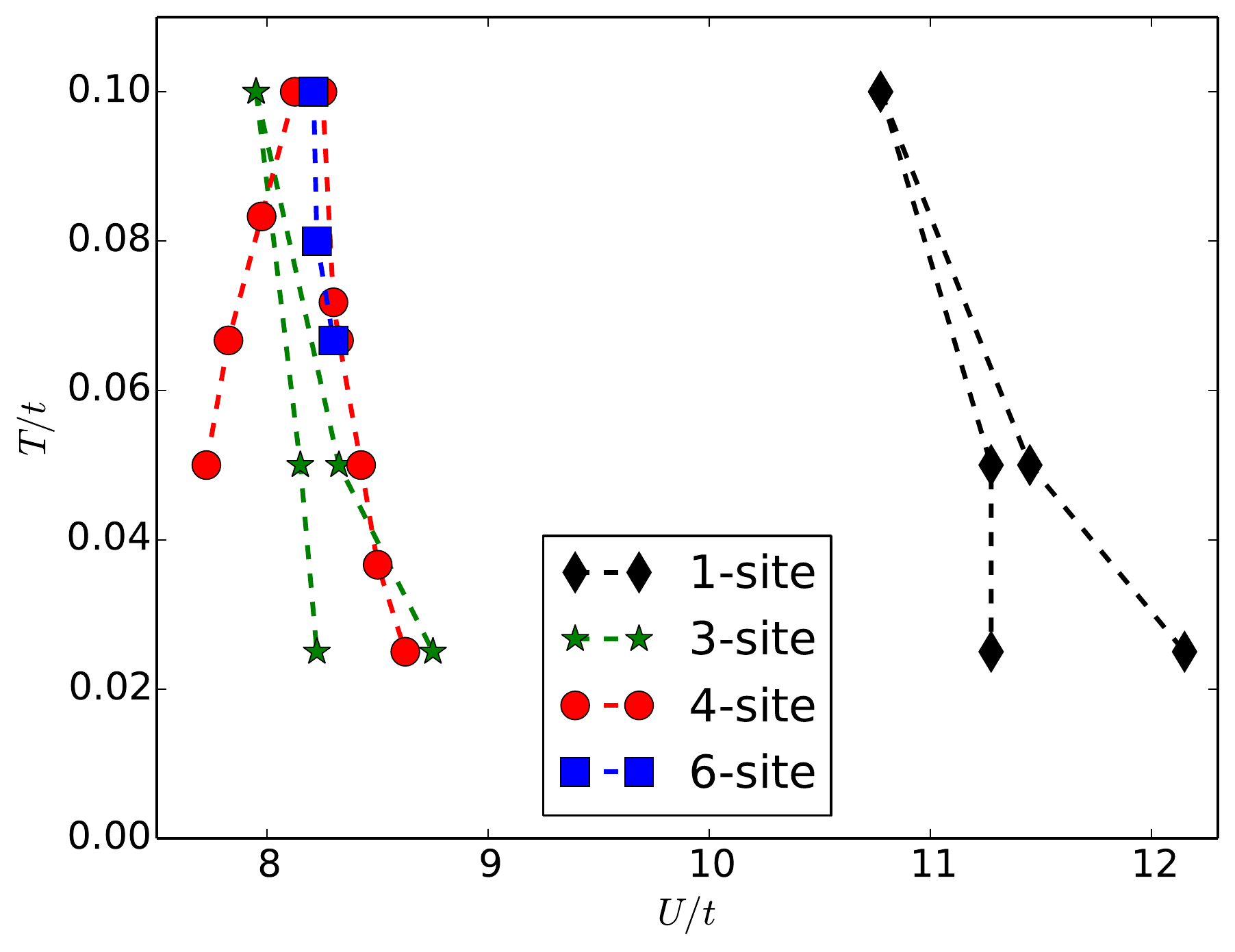}
\caption{\label{fig:phase_diagram_tp1} (Color online) MIT phase boundaries for $t'=t$  from DCA calculations with different impurity cluster sizes ($N_c=1,3,4$ and $6$). For $N_c=6$, only the $U_{c2}(T)$ phase boundary is available.}
\end{figure} 

Considering larger-cluster DCA calculations, in which short-range correlations are accounted for, our $N_c=4$ DCA results exhibit for $t'=0$ an opposite slope for the $U_{c2}(T)$ lines compared to the single-site DMFT result [Fig.~\ref{fig:phase_diagram_1site_4site}(c)] and for $t'=t$ a coexistence region significantly larger than that in DMFT [Fig.~\ref{fig:phase_diagram_1site_4site}(d)]. Comparing the $t'=0$ and $t'=t$ cases, the qualitatively different $U_{c2}(T)$ slopes at $N_c=4$ can be  understood from an increased cluster ground state degeneracy at $t'=t$, as discussed in Sec.~\ref{sec:discussion}. We also notice that the decrease in the $U_c$-range when going from $N_c=1$ to $N_c=4$ DCA for $t'=t$ is weaker than for the  $t'=0$ case. This is in accord with the fact that the perfect AFM nesting in the unfrustrated case is no longer maintained as $t'\to t$.

To arrive at a consistent picture for the triangular lattice in the presence of finite cluster size effects, we collect in Fig.~\ref{fig:phase_diagram_tp1} the MIT phase diagrams for $t'=t$, as obtained for several  different cluster sizes $N_c=1,3,4$ and $6$. We find that going beyond $N_c=1$ DMFT, the phase boundaries converge more rapidly than in the unfrustrated square lattice case. The shapes of the coexistence regime for $N_c=3$ and $N_c=4$ DCA are already similar, while the $U_c$ values converge quickly, e.g. $U_c \sim 8.2t$ at $T=0.1t$ for $N_c=3$, $4$ and $6$. 
This result compares well to several ranges reported for the MIT transition at $t'=t$ in the literature (see for example the Supplementary Material of  Ref.~\onlinecite{Kokalj2013} for a recent summary), and furthermore appears well separated from the estimate of $U/t\approx 10$ for the transition into the magnetically ordered large-$U$ region from Ref.~\onlinecite{Yang2010}. We note that for $N_c=6$ DCA, due to the large computational expense and the sign problem, we cannot go to lower temperatures. However, for the temperatures under investigation, the $U_{c2}$ range appears well converged and in particular the negative slope of  curves are consistently determined. This enhanced convergence shows that, unlike the unfrustrated case, where the AFM fluctuations dominate and open a gap at any $U>0$ \cite{Hirsch1985,Varney2009,Schafer2014}, in the strongly frustrated system, the  magnetic fluctuations compete strongly with other interaction channels, such as charge and pairing, possibly giving rise to non-trivial physics such as a quantum spin liquid state in the insulating side.

\begin{figure}
 \includegraphics[width=\columnwidth]{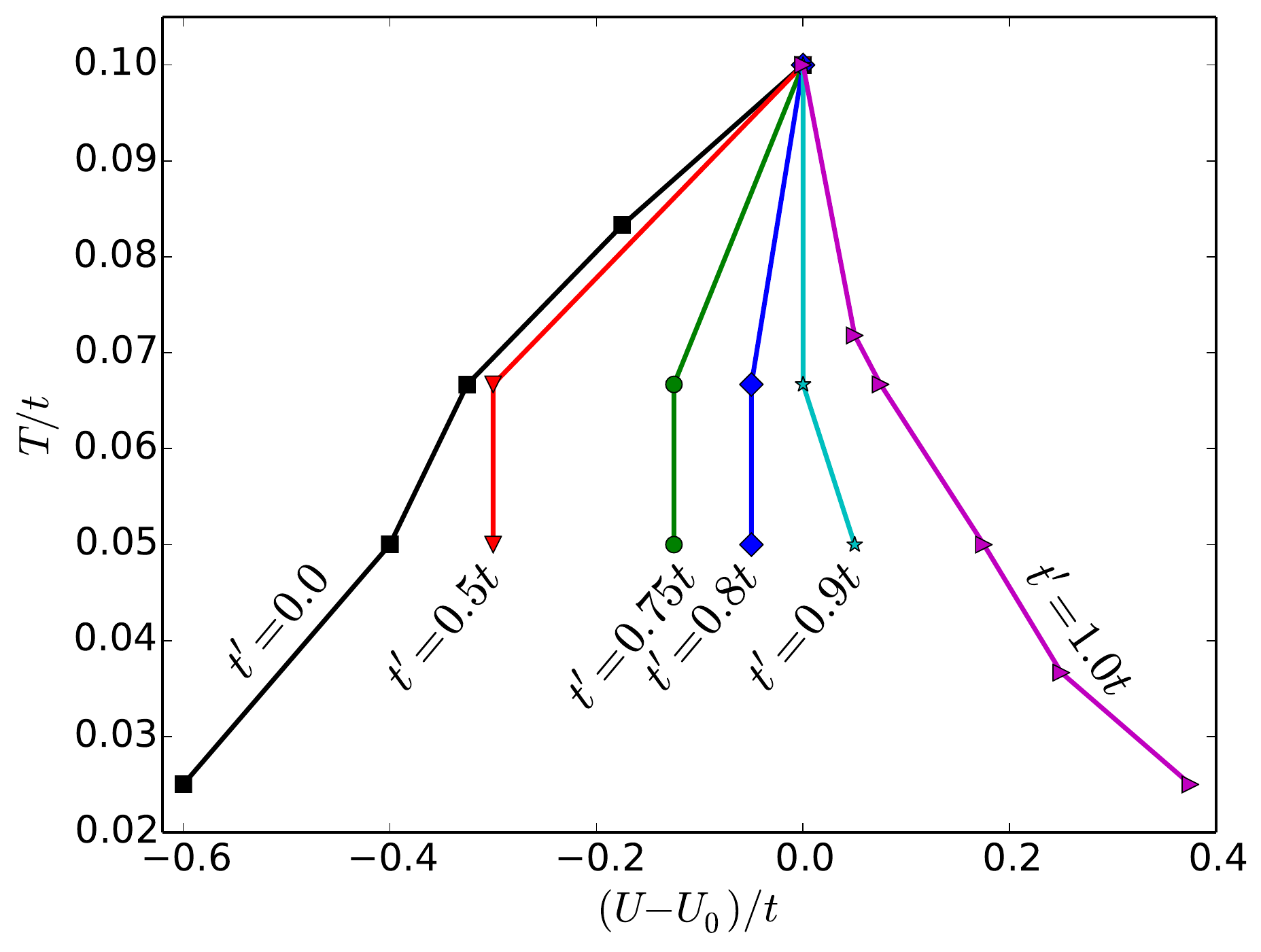}
\caption{\label{fig:phase_diagram_4site} (Color online) Evolution of the $U_{c2}(T)$ phase boundary as  $t'$ increases from $0$ to $t$ within  $N_c=4$ DCA calculations. To ease comparison of the slopes for different values of $t'$, the $U$ values are shifted by $U_0 = U_{c2}(T=0.1t)=5.15t,6.05t,7.25t,7.475t,7.925t$ and $8.25t$ for $t'=0,0.5t,0.75t,0.8t,0.9t$ and $t$, respectively.}
\end{figure} 

Finally, we consider the evolution of the phase boundary upon varying the frustration by tuning $t'$ from $0$ to $t$. Here, we restrict to  $N_c=4$ DCA calculations due to the high computational cost of larger-cluster calculations. Figure~\ref{fig:phase_diagram_4site} shows the evolution of the $U_{c2}(T)$ phase boundary from $N_c=4$ DCA,  when the degree of frustration is increased. The phase boundary has clear positive slope at $t'=0$ and negative slope at $t'=t$. For intermediate values of $t'$, the slope changes gradually and at a critical ratio of $t'_c/t$ between $0.8$ and $0.9$, it becomes nearly vertical. Similar slope changes has been found by Liebsch \textit{et al.}~\cite{Liebsch2009}. The nearly vertical phase boundary was not observed in their calculations at $t'\approx0.8t$, which may be due to the different numerical method (namely $N_c=4$ cellular DMFT with an exact-diagonalization solver) employed in Ref.~\cite{Liebsch2009}. Nevertheless, based on their results at $t'=0.8t$ and $t'=t$, the vertical behavior can appear for $0.8t<t'<t$. 

Even though our calculations are restricted to the  paramagnetic mean-field solution, the negative slope observed in Fig.~\ref{fig:phase_diagram_tp1} as well as the changes of the phase boundary and the vertical $t'_c$ line in Fig.~\ref{fig:phase_diagram_4site} imply important physics. As discussed further below, we expect that as long as the $U_{c2}(T)$ phase boundary exhibits a positive slope (such as at $t'=0$),  AFM fluctuations still render the system unstable towards an AFM Mott insulator. However, when different fluctuations (spin, charge, even pairing) start to compete with each other, the system may enter a more exotic, paramagnetic Mott insulator without spontaneous symmetry breaking, such as a  quantum spin liquid. We anticipate such behavior is related to the change of the slope in the phase boundary line for $t$ beyond $t'_c$. Other works, based on different methods \cite{Tocchio2009,Kyung2006a,Laubach2014,Yamada2014}, also predict a $t'_c$ around $0.8$ to $0.85t$ for the spin liquid phase, which thus would be consistent with our finding.

\section{Understanding the phase diagrams\label{sec:discussion}}

In this section, we fathom several aspects of the $N_c=4$ DCA simulations in order to obtain a more quantitative understanding of the numerical findings.  In particular, we analyze the exact eigenstates of a $N_c=4$ real-space periodic cluster (without bath), the corresponding CT-HYB eigenstate visiting probabilities within the DCA simulations, and estimate the entropy and the (free) energy of the DCA system.

\subsection{Exact diagonalization of the impurity cluster}

\begin{figure}
 \includegraphics[width=\columnwidth]{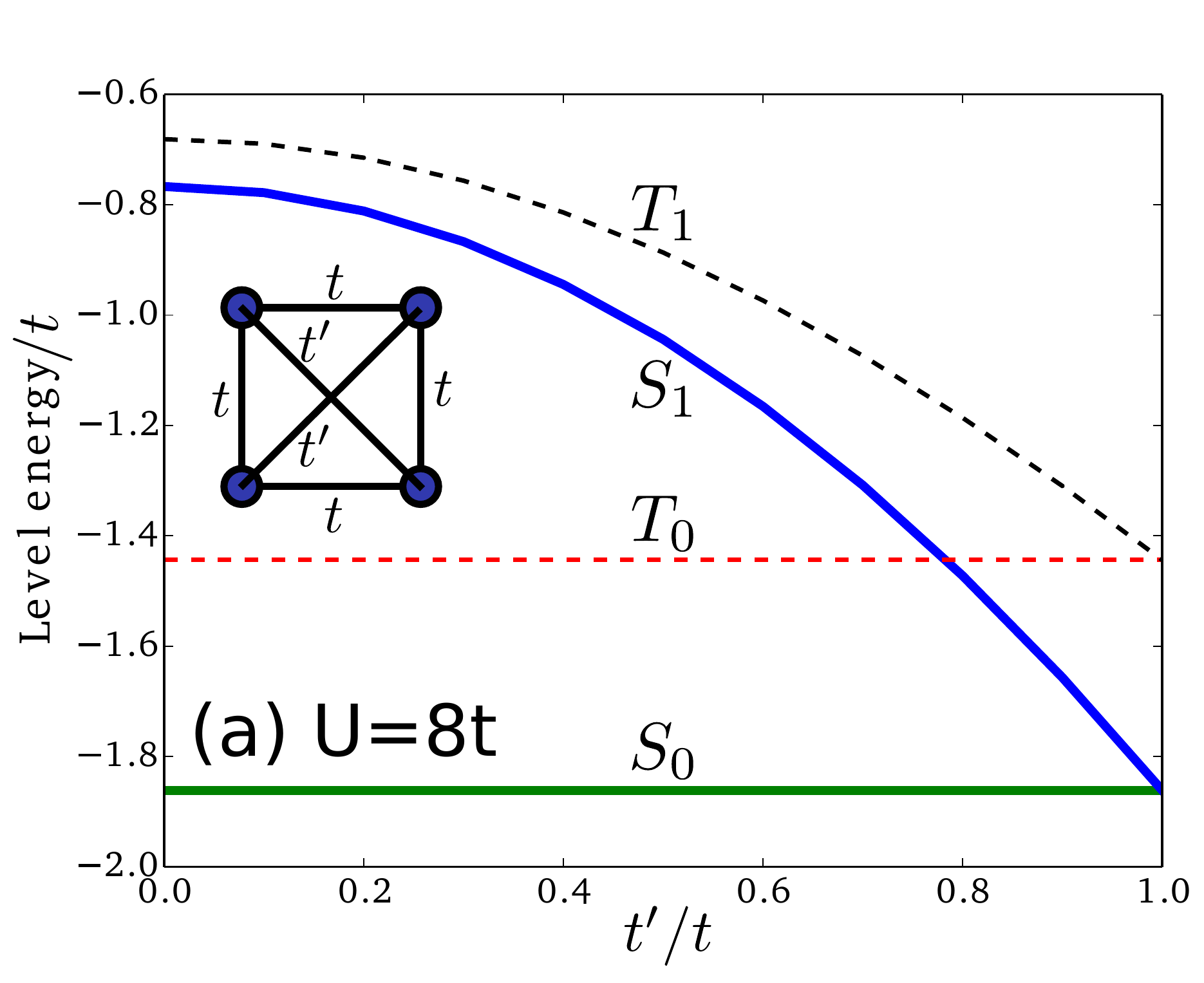}\\
 \includegraphics[width=\columnwidth]{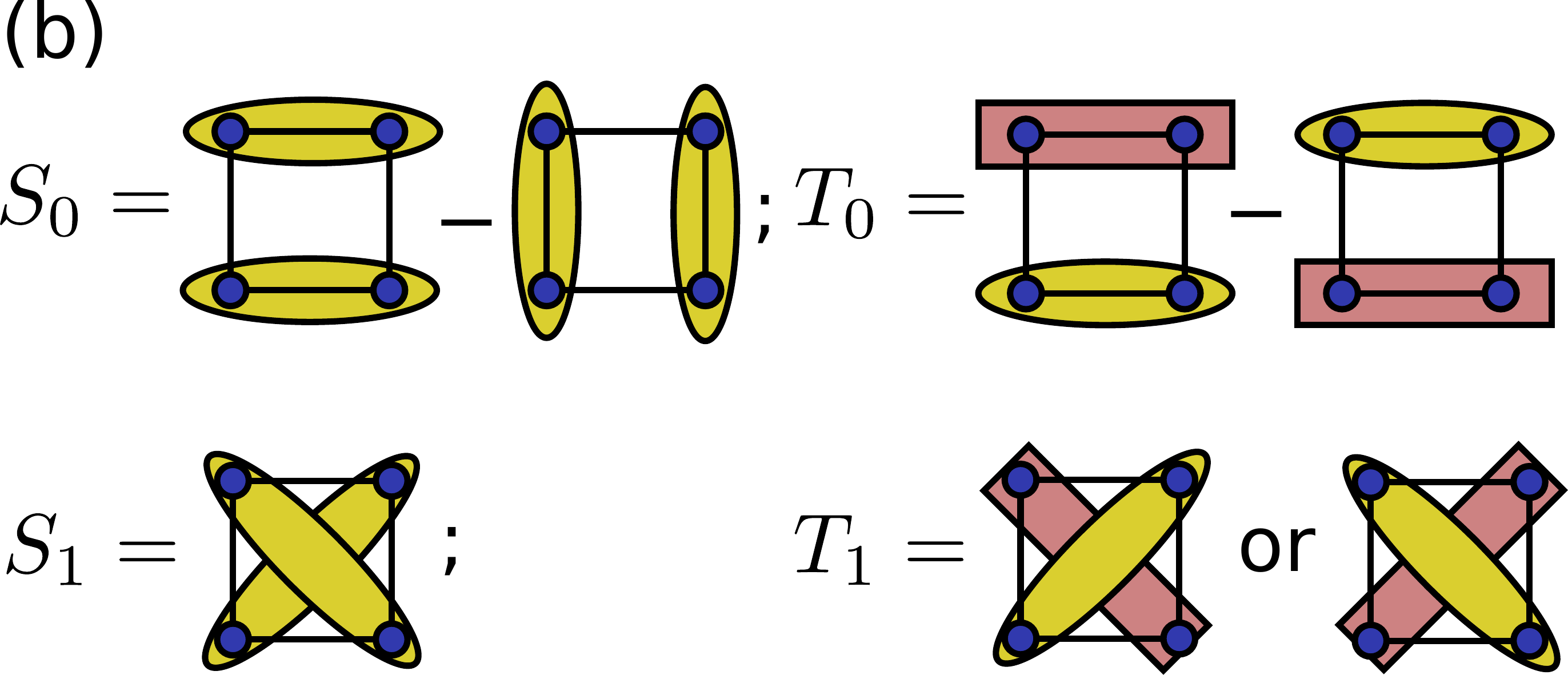}
\caption{\label{fig:ed_4site} (Color online) (a) Exact diagonalization results for the periodic  $N_c=4$ cluster for the evolution of the four lowest-energy levels as $t'$ increases from $0$ to $t$ at half-filling and $U=8t$. The inset shows the real-space cluster with the corresponding hopping amplitudes indicated. (b) Illustration of the four lowest energy levels $S_0$, $S_1$, $T_0$ and $T_1$. An ellipse denotes a singlet state $(|\uparrow\downarrow\rangle-|\downarrow\uparrow\rangle)/\sqrt{2}$ among the two enclosed sites; a rectangle denotes the set of three degenerate triplet states, $|\uparrow\uparrow\rangle$,$|\downarrow\downarrow\rangle$ and $(|\uparrow\downarrow\rangle+|\downarrow\uparrow\rangle)/\sqrt{2}$. Thus, while the $S_0$ and $S_1$ levels are non-degenerate, the degeneracy is three for the $T_0$ and six for the $T_1$ level.}
\end{figure}
We start by analyzing the exact low-energy eigenstates of the $N_c=4$ impurity cluster, obtained via exact diagonalization. Figure~\ref{fig:ed_4site}(a) shows the dependence of the four lowest energy states on the $t'/t$ ratio at $U=8t$. Since the cluster is periodic, the hoppings are arranged equivalent to a $2\times2$ square lattice with diagonal frustration [cf. the inset of Fig.~\ref{fig:ed_4site} (a)]. The low-energy states are illustrated in Fig.~\ref{fig:ed_4site} (b): the ground state $S_0$ is composed out of a resonance between states with two singlets forming along nearest neighbor sites. The  excited level $T_0$ has three degenerate states, corresponding to a triplet along with a singlet connecting nearest neighbor sites. As shown in Fig.~\ref{fig:ed_4site} (a), $S_0$ and $T_0$ are the two lowest energy states over a wide range of $t'$ ($0\leq t'< 0.8$). The energy levels $S_1$ and $T_1$, on the other hand, consists of  singlet and triplet states involving next-nearest neighbor sites with hopping $t'$. The $S_1$ level reaches closer to the ground state only when $t'/t$ increases, in particular beyond $0.8$. 
The behavior is unchanged if $U$ is varied within the relevant range for the MIT. We also emphasize that Fig.~\ref{fig:ed_4site} is used here to obtain a simple framework for the evolution of the MIT phase boundary as $t'$ increases [see Fig.~\ref{fig:phase_diagram_4site}], i.e. at intermediate coupling strength. 
A seemingly similar level spectroscopy was employed in Ref.~\onlinecite{Powell2011} to explore the effects of frustration on the triangular lattice in the Heisenberg limit, corresponding to the large-$U$ regime.

This rather simple analysis turns out to be useful for understanding the behavior of the MIT phase diagrams obtained from the $N_c=4$ DCA simulations. Increasing the $t'/t$ ratio enhances the number of states at low energy: the ground state manifold changes from a single singlet to two-fold degenerate singlet states, and also the first excited level turns from three- to nine-fold degenerate triplet states when varying $t'$ from $0$ to $t$. The higher degeneracy leads to a larger entropy of  the insulating state at $t'=t$ than at $t'=0$, and hence the insulating state is more stable at high temperature~\cite{Laubach2014}. This explains the negative slope of the $U_{c2}(T)$ MIT phase boundary; namely, as temperature decreases, the system goes through a MIT from an insulating to a metallic state in the fully frustrated system, whereas in the unfrustrated case, it is the opposite. Moreover, the changes in the singlet configurations in the ground state manifold suggests that the AFM ground state [captured by the state $S_0$ in Fig.~\ref{fig:ed_4site} (b)] of the  unfrustrated case will be increasingly competed by the $S_1$ state, which instead exhibits diagonal singlets connecting next-nearest neighbor sites. 

\subsection{CT-HYB statistics}
Besides the capability of going to low temperatures and large interaction strength, a further advantage of CT-HYB cluster solver in DCA calculations is the possibility to measure the visiting probabilities of the cluster eigenstates during the QMC simulations~\cite{Werner2006a}. Analyzing such data provides valuable insight into the system's behavior at the MIT.

\begin{figure}
 \includegraphics[width=\columnwidth]{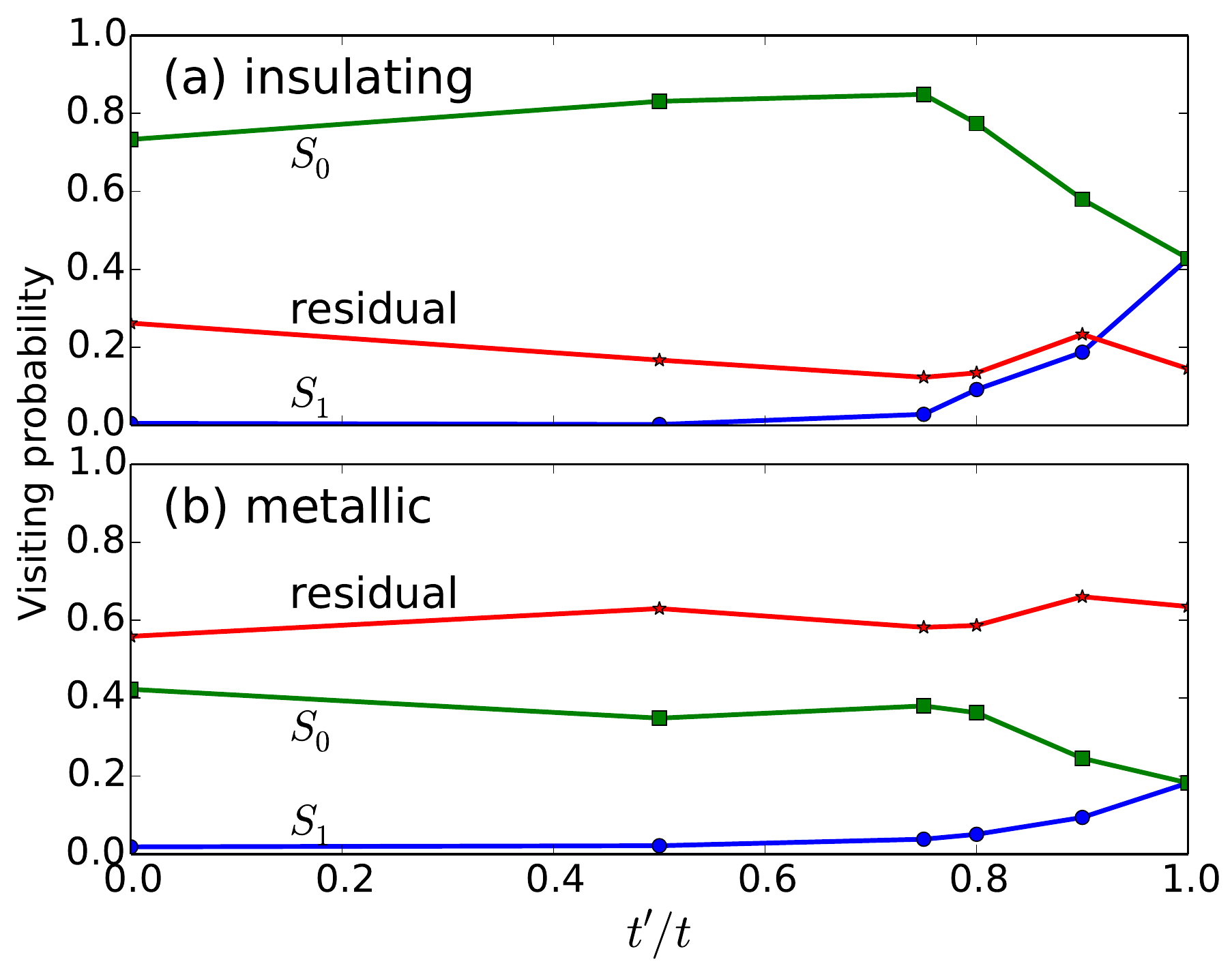}
\caption{\label{fig:eigen_statistics} (Color online) Dependence of the cluster eigenstates visiting probabilities on $t'/t$. The $S_0$ and $S_1$ states are illustrated in Fig.~\ref{fig:ed_4site} (b), while the residual weight is defined as the the sum of the visiting probabilities for all remaining states. The data shown are taken from calculations at $T=0.05t$, with $U$ chosen at the insulating [panel (a)] or metallic [panel (b)] states closest to the $U_{c2}(T)$ phase boundary. The corresponding pairs of $U/t$ in use are $(4.6,4.8)$, $(5.7,5.8)$, $(7.1,7.2)$, $(7.4,7.45)$, $(7.95,8.0)$ and $(8.4,8.45)$ for $t'=0,0.5t, 0.75t, 0.8t, 0.9t$ and $t$, respectively.}
\end{figure}

Based on the exact diagonalization results in Fig.~\ref{fig:ed_4site}, we measured the visiting probabilities for each cluster eigenstate at the metallic and insulating states closest to the $U_{c2}(T)$ phase boundary for the $N_c=4$ DCA simulations. Figure~\ref{fig:eigen_statistics} shows the dependence of the eigenstate visiting probabilities on the $t'/t$ ratio at $T=0.05t$. 
We find that $S_0$ is the most visited state for all values of $t'$. The visiting probability of $S_1$ on the other hand vanishes for $t'=0$ but increases and eventually takes on the same value as for $S_0$ in the fully frustrated case. The other high-energy states (the triplets $T_0$, $T_1$ and all other states) have smaller visiting probabilities, and we thus plot only the sum of all remaining states as the ``residual'' curve in Fig.~\ref{fig:eigen_statistics}. There are two $t'$ values which deserve further attention in Fig.~\ref{fig:eigen_statistics}: (i) beyond $t'\approx0.8t$, the weight for the state $S_1$ starts increasing, marking the onset of the role of frustration, (ii) for the insulating solutions [Fig.~\ref{fig:eigen_statistics}(a)], the state $S_1$ exceeds the residual weight beyond $t'\approx0.9t$. A value of $t'/t$ between $0.8$ and $0.9$ determining the importance of frustration is consistent with $t'_c\approx0.8t$ for the change of the phase boundary slope [cf. Fig.~\ref{fig:phase_diagram_4site}]. We thus believe that this range contains the critical $t'$ value beyond which the AFM order is suppressed, allowing for a paramagnetic Mott insulating phase \cite{Tocchio2009,Kyung2006a,Laubach2014,Yamada2014}.

\subsection{Entropy and free energy}
To further understand the MIT phase boundary slope, we also calculate the entropy and the corresponding free energy of the system. The entropy is obtained from Eq.~\eqref{eq:entropy}, which requires the temperature dependence of the total energy and hence is computationally rather expensive. A less accurate approach is to approximate the entropy of the system by the impurity cluster entropy, which is calculated as $S_{imp}=-\mathrm{Tr}(\hat\rho_{imp}\ln\hat\rho_{imp})$, where $\hat\rho_{imp}$ denotes the density matrix of the impurity cluster and is easily accessed in the CT-HYB solver. 

\begin{figure}
 \includegraphics[width=\columnwidth]{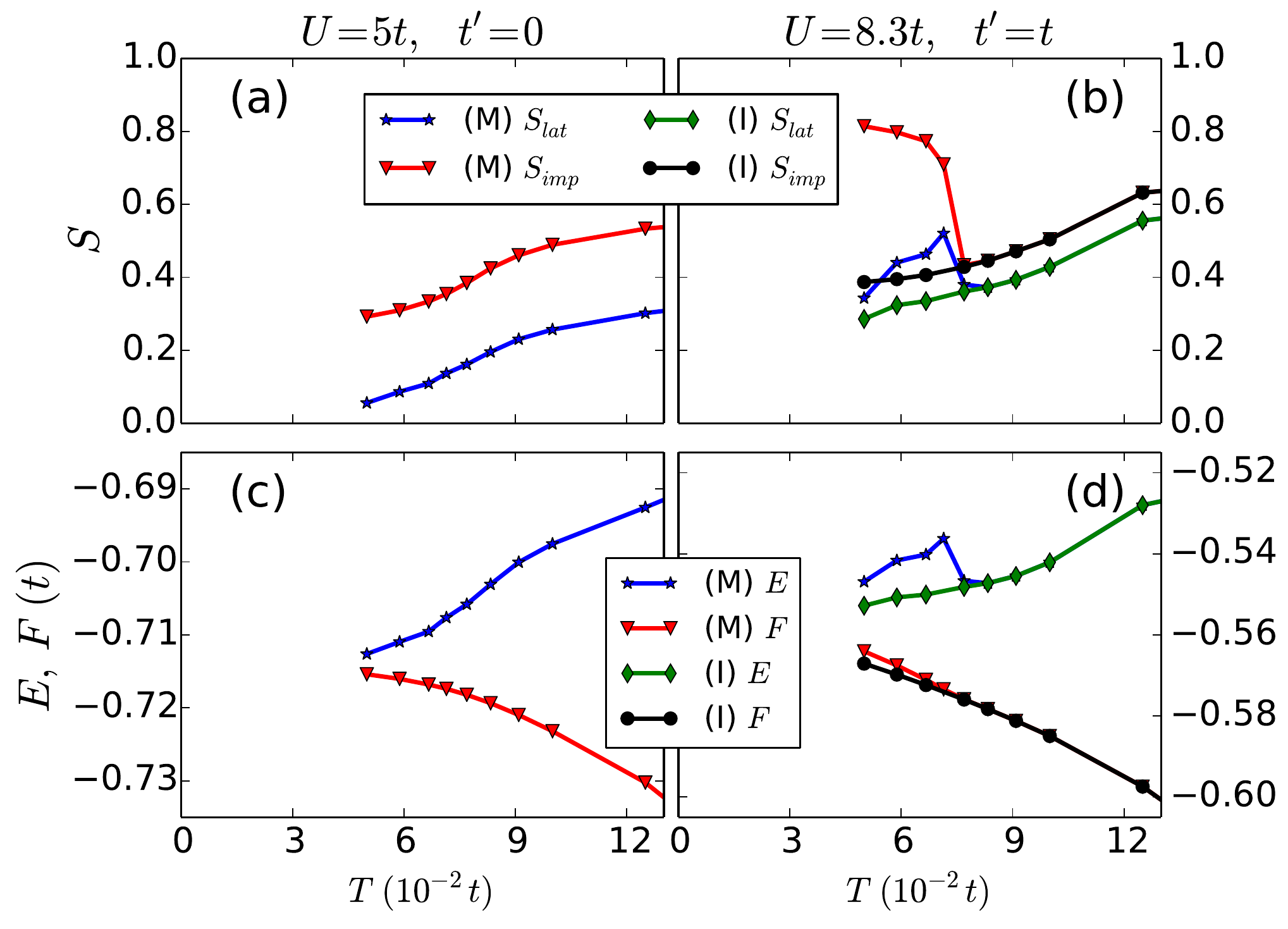}
\caption{\label{fig:energy_entropy} (Color online) Temperature dependence of the entropy $S$, total energy $E$,  and the free energy $F$ per lattice site. The left (right) column corresponds to $t'=0$ ($t'=t$). Panels (a) and (b) show the lattice entropy $S_{lat}$ and the impurity entropy $S_{imp}$. Panels (c) and (d) show the total energy and the free energy. The labels (M) and (I) denote the initial conditions as metallic and insulating states, respectively. Each pair of panels (a,b) and (c,d) shares their legends.}
\end{figure}

We show in Fig.~\ref{fig:energy_entropy}(a), (b) both types of entropy as functions of temperature at $t'=0$ and $t'=t$. For the unfrustrated case [panel (a)], there are only results available for the metallic phase because there is almost no coexistence region at $U=5t$ [Fig.~\ref{fig:phase_diagram_1site_4site}(c)]. However, there are both metallic and insulating curves available for the $t'=t$ case [panel (b)] due to the wider coexistence region. The impurity entropy $S_{imp}$ is seen to overestimate the lattice entropy by a constant shift, except for the metallic case at low temperatures for $t'=t$, where a sizable jump appears in $S_{imp}$. Nevertheless, over a wide  range of parameters, the impurity entropy can be usefully employed for a (fast) estimation of the lattice entropy, if the constant shift is known.

The observed upturn in both the total energy $E$ as well as the entropy $S$, when the temperature $T$ decreases through the MIT critical temperature $T_c\approx 0.08t$ for the case of $t'=t$ [Figs.~\ref{fig:energy_entropy}(b) and (d)] for a metallic initial state, indicates that the final metallic state in this case may be metastable. Note that while the entropy can be shifted by a constant, depending on the large-$T$ reference value (see the discussion in Sec.~\ref{sec:model_methods}.C), the free energy ($F$) difference is unchanged by this constant shift. Thus
the (slightly) lower free energy $F$ seen in Fig.~\ref{fig:energy_entropy}(d) of the insulating state suggests the ground state in this regime to be more likely insulating. Therefore, the first order transition line inside the coexistence regime may be located well to the left of the  $U_{c2}(T)$ curve.

\section{Conclusions\label{sec:conclusions}}
We have investigated the Mott MIT for the Hubbard model on the triangular lattice using DCA quantum Monte Carlo simulations. Controlling the degree of frustration by varying the hopping ratio $t'/t$, we studied the evolution of the MIT phase boundary as the geometric frustration increases. Several cluster sizes have been used to convey a consistent picture. We furthermore analyzed the changes of the phase boundary from several perspectives: in the periodic cluster limit, by measuring the visiting probabilities of cluster eigenstates in the QMC simulations, and based on entropy and free energy considerations.

Several differences in the MIT phase diagrams between the unfrustrated ($t'=0$) and fully frustrated ($t'=t$) cases are exposed. A central difference is the change of the $U_{c2}(T)$ phase boundary slope from positive at $t'=0$ to negative at $t'=t$. From our results for  various values of the ratio $t'/t$, we estimated a critical value $t'_c$ between $0.8t$ and $0.9t$ for which the phase boundary is close to vertical. By analyzing the cluster eigenstates and the QMC statistics, $t'_c$ relates to the formation of the $S_1$ singlet along the direction of the hopping $t'$, which eventually becomes degenerate with the $S_0$ singlet along the hopping  $t$ direction. The increase in the ground state degeneracy in the frustrated case weakens  antiferromagnetic fluctuations considerably when $t'> t'_c$, such that various short-range fluctuations compete. In essence, this may lead to a paramagnetic Mott insulator, which is possibly a quantum spin liquid state, as suggested in Refs.~\onlinecite{Tocchio2009,Kyung2006a,Laubach2014,Yamada2014}. On the other hand, the total energy and entropy results for the triangular lattice case suggest that the first order MIT phase boundary is located away from the DCA estimated $U_{c2}(T)$ line.

Implications of these results for the physics of frustrated electronic systems would be that
observing such an anticipated change in the phase boundary slope can be a signal for the appearance of exotic Mott insulating phases. Several experimental phase diagrams for organic materials \cite{Limelette03,Kanoda2011,Kurosaki2005} show that the MIT phase boundary varies in a similar manner as our results when frustration is enhanced.

To improve the numerical results, we suggest to employ larger-cluster DMFT calculations with cluster sizes being multiples of three such  that the triangular lattice frustration is maintained explicitly. Therefore, cellular DMFT~\cite{Kotliar2001} may be a more appropriate scheme, if one could find ways to overcome the minus-sign problem that plagues large-cluster-size and low-temperature simulations. In addition, calculations with symmetry breaking mean-field baths would be  useful for specifying magnetic transitions and could provide a more direct prediction for the location of the loss of magnetic order in the insulating state.

\section*{Acknowledgments}
We thank V. Dobrosavljevic and K. Kanoda for valuable discussions on the nature of the Mott transition and organic materials and also acknowledge discussions with A. Georges, E. Gull, A. Liebsch, A. J. Millis, T. Pruschke and Chuck-Hou Yee.
X.Y.X. and Z.Y.M. are supported by the National Thousand-Young-Talents Program of China, their computations were preformed on TianHe-1A, the National Supercomputer Center in Tianjin, China. K.S.C. is supported by the Deutsche Forschungsgemeinschaft (DFG) under the grant number AS120/8-2 (Forschergruppe FOR 1346). H.T.D. and S.W. acknowledge support from the DFG within projects FOR 1807, RTG 1995, and WE3649/3-1, as well as the allocation of computing time at J\"ulich Supercomputing Centre and RWTH Aachen University through JARA-HPC.  
We used the code for the CT-HYB solvers~\cite{Werner2006} from the TRIQS project~\cite{triqs_project}, from ALPS-2.0~\cite{Hafermann2013,ALPS2.0} and the one written by P. Werner and E. Gull, based on the ALPS-1.3 library~\cite{ALPS1.0}.

\appendix
\section{The sign problem\label{app:sign_problem}}
In QMC simulations of fermionic systems, the sign problem is the main challenge that prohibits the study of large-scale systems and low temperatures. In fact, the sign problem belongs to the class of NP-hard problems~\cite{Troyer2005}. Swapping two fermions causes a change of sign in the overall wavefunction, thus when constructing the QMC probability distribution, it is possible that the probability is not positive definite, which can lead to measurements with large statistical errors~\cite{Troyer2005,Iazzi2014}. In many cases, the average sign of the probability distribution, $\langle \mathrm{sgn} \rangle$, decreases exponentially as the inverse temperature $\beta=1/T$ or the system size increases~\cite{Troyer2005, Gull2011}, causing severely large errors in QMC measurements.

\begin{figure}[t]
 \includegraphics[width=\columnwidth]{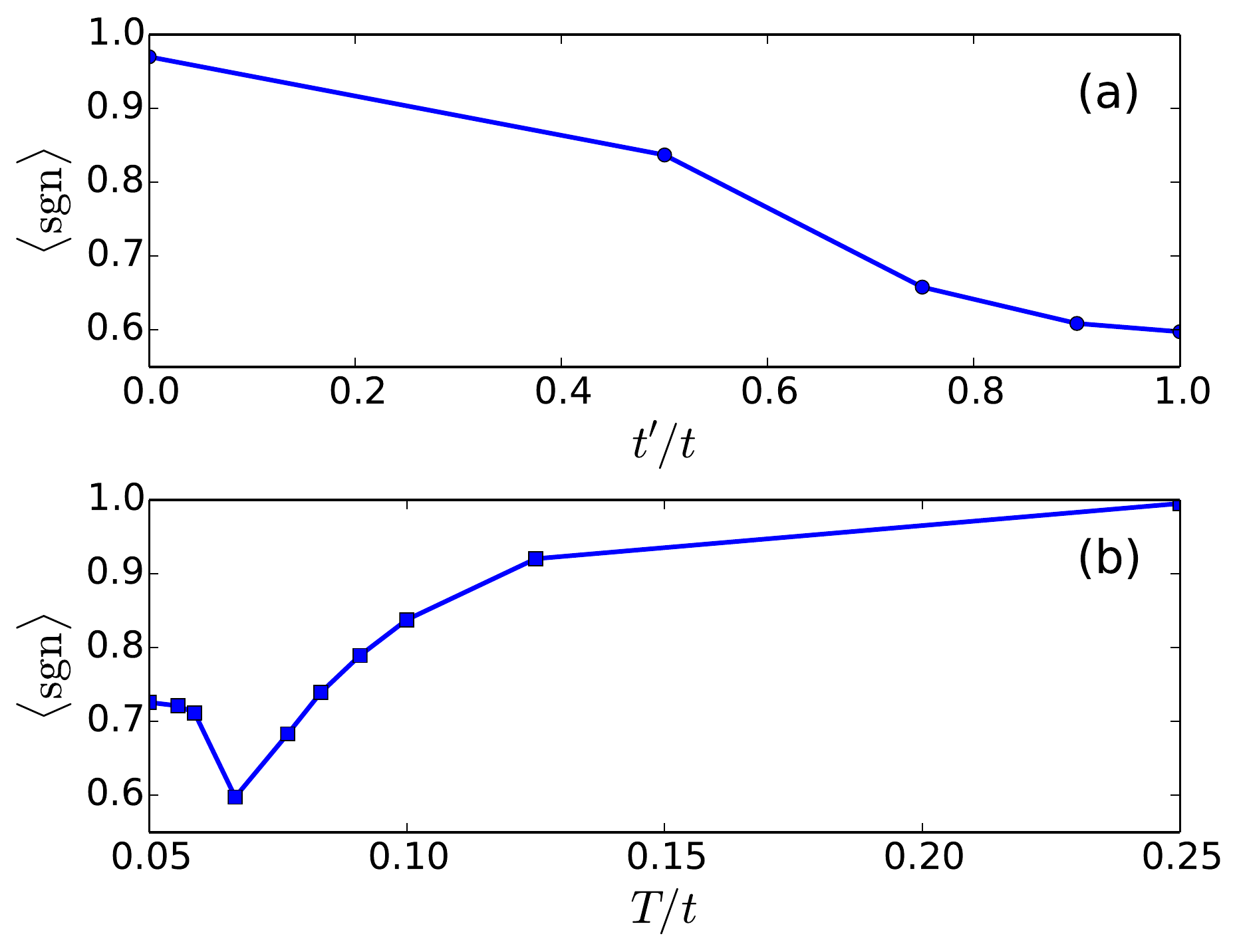}
\caption{\label{fig:sign} (a) Dependence of the average sign, $\langle \mathrm{sgn} \rangle$, on the degree of frustration, the $t'/t$ ratio, at $T=0.067t$ for $U$ values chosen  close to the phase boundary on the insulating side. (b) Temperature dependence of $\langle \mathrm{sgn} \rangle$ for $t'=t$ and $U=8.35t$.}
\end{figure} 

While DMFT is an appropriate approximation to avoid the sign problem, the issue cannot be completely resolved; in some cases such as for frustrated systems examined within this study, the sign problem in the impurity solver becomes severe. Therefore for $N_c=4$ DCA calculations using the CT-HYB solver, our temperatures are restricted above $T=0.025t$ for the fully frustrated case ($t=t'$). Since the complexity of the CT-HYB solver increases exponentially  with the cluster size, given limited computational resources, it is not possible to extend this solver to larger cluster sizes. The CT-INT solver was used instead for the $N_c=6$ DCA calculations, which is exposed more to the sign problem, thus in these DCA calculations we only considered temperatures larger than $T=0.07t$.

To quantify the sign problem in this study, we show in Fig.~\ref{fig:sign} the average sign, $\langle \mathrm{sgn} \rangle$, that we observe in our $N_c=4$ DCA simulations. At intermediate temperatures, $T=0.067t$, the average sign decreases as the $t'/t$ ratio increases towards $1$ [panel (a)]. Frustration is the main cause for the decrease of $\langle \mathrm{sgn} \rangle$, as a higher degree of frustration implies a larger degeneracy on the impurity cluster, thus increasing the chance for swapping impurity electrons in the simulations. Figure~\ref{fig:sign}(b) shows the temperature dependence of the sign for the isotropic case ($t'=t$). Interestingly, $\langle \mathrm{sgn} \rangle$ is seen to decrease along with temperature only for $T\le0.067t$, while it \textit{increases} below this temperature. This however does not contradict the fact that low temperature worsens the average sign; instead this dip-feature signals the MIT,  which occurs at $T\sim 0.067t$. The insulating state exhibits a smaller sign due to its closeness to the atomic limit, and the visiting probability is distributed over fewer impurity eigenstates, such that there is a larger chance for sign changes in the  QMC probability distribution.

\section{Details of the entropy calculations\label{app:entropy}}

To estimate the entropy within the DMFT framework, one can employ the fundamental thermodynamic relation~\cite{Georges1996,Werner2005,Mikelsons2009,LeBlanc2013},
\begin{equation}
 S(T) = \int_0^T \dfrac{C_v}{T'} dT' = S(T_{ref}) - \int_T^{T_{ref}} \dfrac{C_v(T')}{T'} dT'.
\end{equation}
Since $C_v(T) = d E/dT$ (where $E$ is the total energy), using $\beta=1/T$, and after integrating by parts, the following formula for the entropy is obtained:
\begin{equation}
 S(\beta) = S(\beta_{ref}) + \beta E(\beta)|_{\beta{ref}}^\beta - \int_{\beta_{ref}}^\beta E(\beta')d\beta',
\end{equation}
where $\beta_{ref} = 1/T_{ref}$. For the reference temperature, $T_{ref}\to\infty$ (or $\beta_{ref} = 0$) is usually taken~\cite{Georges1996,Werner2005,Mikelsons2009}, 
since $S(T_{ref}\to\infty)=\ln 4$ is known exactly. In this way, Eq.~\eqref{eq:entropy} is obtained. 

\begin{figure}[h]
 \includegraphics[width=\columnwidth]{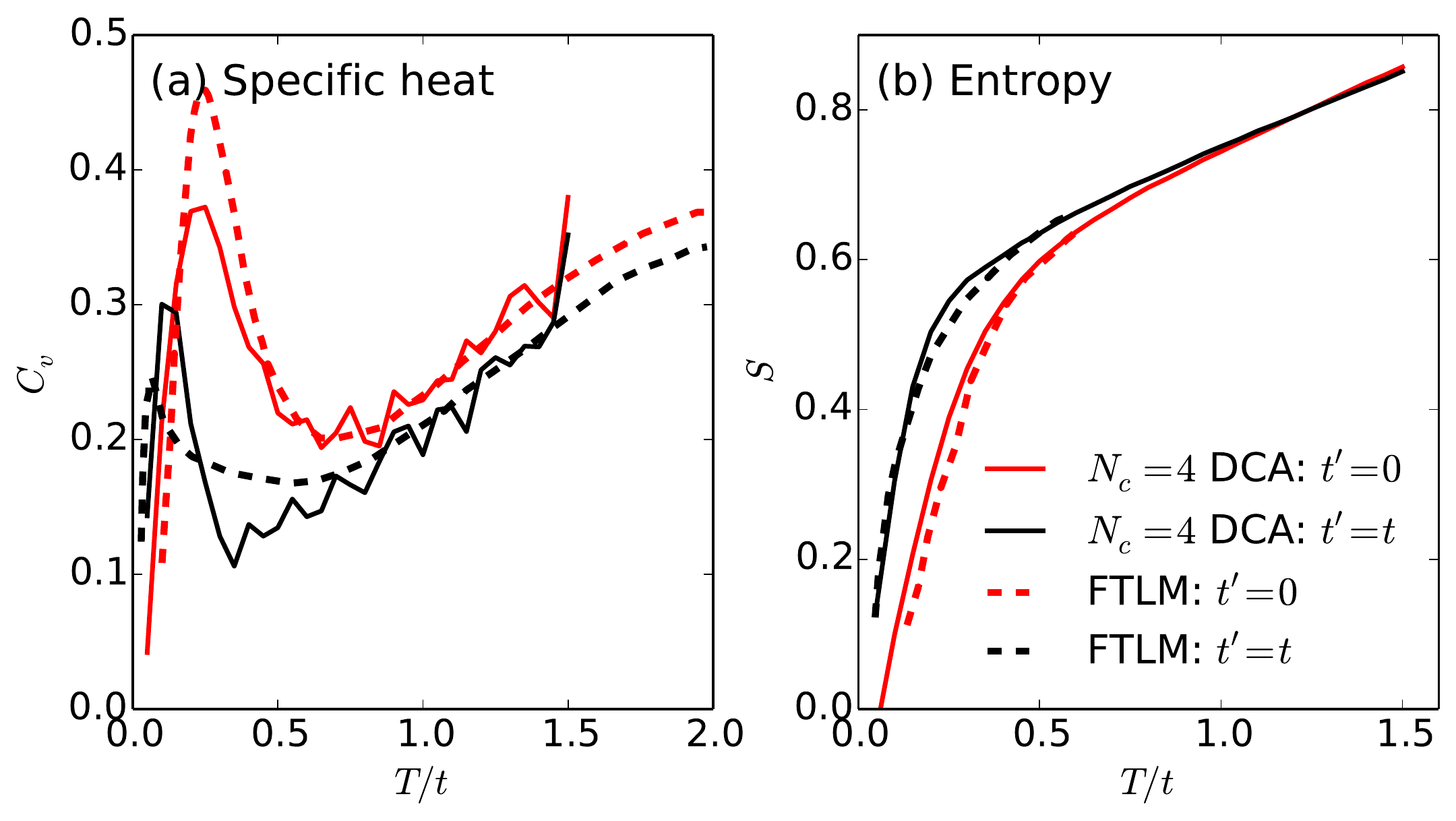}
\caption{\label{fig:compare_cv_s} (Color online) Comparison between $N_c=4$ DCA calculations (solid lines) and finite temperature Lanczos method (FTLM) results~\cite{Kokalj2013} (dashed lines) for (a) the specific heat and (b) the entropy per 
site. Two cases 
$t'=0$ (light - red online) and $t'=t$ (dark - black online) are shown for $U=10t$. The FTLM data are extracted from the Supplementary Material of Ref.~\onlinecite{Kokalj2013}.}
\end{figure} 

However, the drawback of using $T_{ref}\to\infty$ as the reference temperature of the thermodynamic integration is the need for extended calculations up to rather high temperatures. Using instead a reference temperature that is located closer to  the temperature range of interest, can provide more accurate results. We demonstrate this point and verify our entropy calculations by comparing  with results from finite temperature Lanczos method (FTLM) calculations [see Ref.~\onlinecite{Kokalj2013} and references therein]. In particular, we use FTLM data taken at $T_{ref}=0.6t$ from Ref.~\onlinecite{Kokalj2013} as a reference temperature point and compare our DCA calculations at $U=10t$ for both the unfrustrated ($t'=0$) and the fully frustrated ($t'=t$) cases with the FTLM results provided in Ref.~\onlinecite{Kokalj2013}. Figure~\ref{fig:compare_cv_s} shows the entropy and specific heat, which are very similar for the two methods. 
In particular, the low-temperature peaks for the specific heat are closely located and the same trends are observed as $t'$ increases from $0$ to $t$. 
These results confirm the validity of our entropy calculations within the DCA method. However, good reference point data for the entropy are not available for the cases that we consider in the main text, thus $T_{ref}\to\infty$ is used as the reference temperature. We verified that the difference in entropy due to the reference point is well-approximated by a constant shift. Therefore, for our entropy calculations in Sec.~\ref{sec:discussion}.~C, where the central goal is the comparison of the free energy {\it differences} between the two DCA solutions, the choice of $T_{ref}\to\infty$ provides an acceptable reference point.

\bibliographystyle{apsrev4-1}
\bibliography{triangle}
\end{document}